\documentclass[%
 aip,
 amsmath,amssymb,
preprint,%
fleqn,
]{revtex4-1}

\usepackage{graphicx}
\usepackage{dcolumn}
\usepackage{bm}

\begin{document}


\title{ Existence and stability of dust ion acoustic double layers described by the combined MKP-KP equation}

\author{Sankirtan Sardar}
\affiliation{ Department of Mathematics, Jadavpur University, Kolkata - 700032, India.}%
\author{Anup Bandyopadhyay}
\email{abandyopadhyay1965@gmail.com}%
\affiliation{ Department of Mathematics, Jadavpur University, Kolkata - 700032, India.}%
\author{K. P. Das}
\affiliation{ Department of Applied Mathematics, University of Calcutta, 92 Acharya Prafulla Chandra Road, Kolkata - 700009, India.}%
\begin{abstract}
\noindent The purpose of this paper is to expand the recent work of Sardar \textit{et al.} [Phys. Plasmas \textbf{23}, 123706 (2016)] on the existence and stability of alternative dust ion acoustic solitary wave solution of the combined  modified Kadomtsev Petviashvili - Kadomtsev Petviashvili (MKP-KP) equation in a nonthermal plasma. Sardar \textit{et al.} [Phys. Plasmas \textbf{23}, 123706 (2016)] have derived a combined MKP-KP equation to describe the nonlinear behaviour of the dust ion acoustic wave when the coefficient of the nonlinear term of the KP equation tends to zero. Sardar \textit{et al.} [Phys. Plasmas \textbf{23}, 123706 (2016)] have used this combined MKP-KP equation to investigate the existence and stability of the alternative solitary wave solution having a profile different from $\mbox{sech}^{2}$ or  $\mbox{sech}$ when $L>0$, where $L$ is a function of the parameters of the present plasma system. In the present paper, we have considered the same combined MKP-KP equation to study the existence and stability of the double layer solution and it is shown that double layer solution of this combined MKP-KP equation exists if $L=0$. Finally, the lowest order stability of the double layer solution of this combined MKP-KP equation has been investigated with the help of multiple scale perturbation expansion method of Allen and Rowlands [ J. Plasma Phys. \textbf{50}, 413 (1993)]. It is found that the double layer solution of the combined MKP-KP equation is stable at the lowest order of the wave number for long-wavelength plane-wave perturbation.
\end{abstract}
\maketitle

\section{Introduction}
In last few decades, much emphasis has been given to understand the nonlinear wave processes in dusty plasmas. Formation of double layers (DLs) in a multispecies unmagnetized or magnetized plasma is an interesting nonlinear phenomena in many areas of space plasma physics \cite{block1972potential,mozer1977observations,sato1980ion,hudson1981electrostatic,temerin1982observations,schamel1986electron,kim1987theory,schamel1987stability,bostrom1988characteristics,bostrom1992observations,han1994weak,dovner1994freja,mozer1998direct,ergun2002parallel,main2006double,kim2007modified,angelopoulos2008first,ergun2009observations,li2015characteristics} as well as in various types of laboratory plasmas \cite{coakley1979laboratory,leung1980formation,stenzel1980v,sato1981ultrastrong,sato1982double,sato1983stationary,nakamura1982potential,torven1979observations,torven1980properties,baker1981studies,jovanovic1982three}. The S3-3, Viking, FAST and Polar Satellites, THEMIS spacecrafts  have shown the presence of DLs in many space plasma environments  \cite{mozer1977observations,sato1980ion,mozer1980satellite,hudson1981electrostatic,temerin1982observations,bostrom1988characteristics,
mozer1998direct,ergun2002parallel,angelopoulos2008first,ergun2009observations,li2015characteristics}. Computer simulation also indicates the existence of DLs \cite{sato1980ion,sato1981numerical,temerin1982observations,hudson1981electrostatic,baker1981studies}.

Several authors \cite{bharuthram1986large,bandyopadhyay2000ion,bandyopadhyay2001stability,ghosh2008ion,ourabah2013weakly,alam2014effects,masud2015dust} investigated small or arbitary amplitude ion acoustic (IA)/ Dust ion acoustic (DIA) DLs in different plasma systems with or without magnetic field. For the first time, Bharuthram \& Shukla \cite{bharuthram1986large} have studied the finite amplitude IA double layer (DL) in a collisionless unmagnetized plasma with cold ions and two distinct populations of Boltzmann-Maxwellian distributed electrons. They have used a combined modified Korteweg-de Vries- Korteweg-de Vries (MKdV-KdV) equation to study the propagation characteristics of IA double layers. Bandyopadhyay \& Das \cite{bandyopadhyay2000ion} have derived combined modified Korteweg-de Vries - Zakharov-Kuznetsov (MKdV-KdV-ZK) equation to describe IA double layers in a collisionless magnetized plasma consisting of warm ions and nonthermal electrons. In a later paper, Bandyopadhyay \& Das \cite{bandyopadhyay2001stability} used the multiple-scale perturbation expansion method of Allen and Rowlands \cite{allen1993determination}, to study the DL solution of the combined MKdV-KdV-ZK equation. Ghosh \& Bharuthram \cite{ghosh2008ion} have investigated IA double layers in a collisionless unmagnetized electron-positron-ion-dust plasma (e-p-i-d plasma) with isothermally distributed electrons and positrons. They have derived  a combined MKdV-KdV equations to discuss IA double layers. Ourabah \& Tribeche \cite{ourabah2013weakly} have found that the critical values of the parameters for which DIA double layers exist in a dusty plasma with nonextensive electrons. The analytical and numerical study of small amplitude DIA double layers in a collisionless unmagnetized dusty plasma system consisting of inertial ions, negatively charged immobile dust, and superthermal electrons with two distinct temperatures are investigated by Alam \textit{et al.} \cite{alam2014effects}. Masud \textit{et al.} \cite{masud2015dust} have derived a combined MKdV-KdV equation to discuss the DL solution in a collisionless unmagnetized dusty plasma consisting of negatively charged static dust grains, Boltzmann electrons, and inertial ions. In the present paper, we have used the combined modified Kadomtsev Petviashvili - Kadomtsev Petviashvili (MKP-KP) equation  of Sardar \textit{et al.} \cite{sardar2016existence} to investigate the existence and stability of the DL solution.

 The present problem is an extension of earlier works of Sardar \textit{et al.} \cite{sardar2016stability,sardar2016existence} So, to explain the present problem, we have considered the following points by summarizing the previous works of Sardar \textit{et al.} \cite{sardar2016stability,sardar2016existence} :
\begin{itemize}
	\item Sardar \textit{et al.} \cite{sardar2016stability} have derived a Kadomtsev Petviashvili (KP) equation having nonlinear term $\displaystyle AB_{1} \frac{\partial}{\partial \xi}\Big(\phi^{(1)} \phi^{(1)}_{\xi}\Big)$ to describe the nonlinear behaviour of DIA waves in a collisionless unmagnetized dusty plasma  consisting of  warm adiabatic ions, static negatively charged dust grains, nonthermal electrons and isothermal positrons, where $\phi^{(1)}$ is the first order perturbed electrostatic potential, $\xi$ is the stretched spatial coordinate, $A$ and $B_{1}$ are the functions of the parameters involved in the system. This KP equation can describe the nonlinear dynamics of DIA waves only when $B_{1} \neq 0$ and $B_{1}$ is not close to zero.
	\item When $B_{1}=0$, Sardar \textit{et al.} \cite{sardar2016stability} have derived a modified KP (MKP) equation having nonlinear term $\displaystyle AB_{2} \frac{\partial}{\partial \xi}\Big[\Big(\phi^{(1)}\Big)^{2} \phi^{(1)}_{\xi}\Big]$ to describe the nonlinear behaviour of DIA waves, where $B_{2}$ is also a function of the parameters involved in the system.
	\item When $B_{1} \neq 0$ but $B_{1}$ is close to zero, Sardar \textit{et al.} \cite{sardar2016existence}  have shown that neither KP nor MKP equation can describe the nonlinear behaviour of DIA waves because the amplitude of the solitary wave solution defined by the KP equation assumes a very large numerical value when $B_{1}$ is close to zero.
	\item When $B_{1} \neq 0$ but $B_{1}$ is close to zero, Sardar \textit{et al.} \cite{sardar2016existence} have derived a combined  MKP-KP equation which efficiently describes the nonlinear behaviour of DIA waves.
	\item Sardar \textit{et al.} \cite{sardar2016existence}  have obtained the alternative solitary wave solution of the combined MKP - KP equation having a profile different from $\mbox{sech}^{2/r}$ for any strictly positive real value of $r$. They have also shown that this alternative solitary wave solution is exactly same as the solitary wave solution of the MKP equation if the coefficient of the nonlinear term of the KP equation tends to zero.
	\item Sardar \textit{et al.} \cite{sardar2016existence} have investigated the condition for the existence of the alternative solitary wave solution of the combined MKP-KP equation and they have shown that the alternative solitary wave solution of the combined MKP-KP equation exists if and only if $L>0$, where $L$ is a function of the parameters.
	\item Sardar \textit{et al.} \cite{sardar2016existence} have shown that the alternative solitary wave solution of the combined MKP - KP equation cannot describe the nonlinear behaviour of DIA waves when $L=0$ or $L \approx O(\epsilon)$ and they have reported that further investigation is necessary when $L=0$ or $L \approx O(\epsilon)$, where $\epsilon$ is a small parameter.
	\item In the present paper, we want to find a consistent solution of the combined MKP - KP equation when $L=0$. In fact, we see that one can get a DL solution of the combined MKP - KP equation when $L=0$.
	\item In this paper, we have also investigated the lowest order stability of the DL solution of the combined MKP-KP equation.
\end{itemize}

This paper is organized as follows: The evolution equation is given in \S \ref{P5_Evolution_equation}. In \S \ref{P5_Double_layer_solution}, we have derived the double layer solution of the combined MKP - KP equation. The stability of the double layer solution has been investigated in \S \ref{P5_Stability_Double_layer}. Finally, conclusions are given in \S \ref{P5_conclusions}.

\section{Evolution equation}\label{P5_Evolution_equation}
Starting from the set of basic equations consisting of the equation of continuity of ions, equation of motion of ions, the pressure equation for ion fluid, the Poisson equation, the equation for the number density of the nonthermal electrons of Cairns \textit{et al.} \cite{cairns1995electrostatic}, the equation for the number density of isothermal positrons and the unperturbed charged neutrality condition, Sardar \textit{et al.} \cite{sardar2016existence} have derived following combined  MKP-KP equation:
\begin{equation}\label{P5_Combined_MKP_KP_equation}
\frac{\partial}{\partial \xi}\Big[\phi^{(1)}_{\tau} + AB_{1}
\phi^{(1)} \phi^{(1)}_{\xi}  + AB_{2} (\phi^{(1)})^{2} \phi^{(1)}_{\xi} + \frac{1}{2} A C \phi^{(1)}_{\xi \xi \xi}\Big]
+ \frac{1}{2} AD \Big( \phi^{(1)}_{\eta \eta } + \phi^{(1)}_{\zeta \zeta}\Big)  = 0.
\end{equation}
This equation describes the nonlinear behaviour of DIA waves in a collisionless unmagnetized dusty plasma consisting of warm adiabatic ions,  static negatively charged dust grains, nonthermal electrons and isothermal positrons when $B_{1} \neq 0$ but $B_{1}$ is close to zero. Here $\xi$, $\eta$, $\zeta$ are the stretched spatial coordinates and $\tau$ is the stretched time coordinate. The coefficients $A$, $B_{1}$, $B_{2}$, $C$ and $D$ are, respectively given by the  following equations:
\begin{equation}\label{P5_expressin of A}
A = \frac{1}{1-p}\frac{(M_{s}^{2} V^{2}- \gamma\sigma_{ie})^{2}}{V M_{s}^{2}},
\end{equation}
\begin{equation}\label{P5_expression of B_{1}}
B_{1} = \frac{1}{2} \Big[(1- p)\frac{3 M_{s}^{2} V^{2}+ \gamma (\gamma-2)\sigma_{ie}}{(M_{s}^{2} V^{2}- \gamma \sigma_{ie})^{3}}-\Big(\mu - \frac{p}{\sigma_{pe}^2}\Big)\Big],
\end{equation}
\begin{eqnarray}\label{P5_expression of B_{2}}
B_{2} = \frac{1-p}{4(M_{s}^{2} V^{2}- \gamma \sigma_{ie})^{5}}\Big[15 M_{s}^{4} V^{4} &+& \gamma(\gamma^{2}+13\gamma-18) M_{s}^{2} V^{2}\sigma_{ie} \nonumber \\
&+& \gamma^{2}(\gamma-2)(2\gamma-3) \sigma_{ie}^{2}\Big]-\frac{3}{2}Q_{3},
\end{eqnarray}
\begin{equation}\label{P5_expression of C}
C=\frac{1-p}{M_{s}^{2}-\gamma \sigma_{ie}},
\end{equation}
\begin{equation}\label{P5_expression of D}
D = (1-p) \frac{M_{s}^{2} V^{2}}{(M_{s}^{2} V^{2}-\gamma \sigma_{ie})^{2}},
\end{equation}
and the constant $V$ is determined by
\begin{equation}\label{P5_expression of V}
V^{2}=1,
\end{equation}
where $M_{s}$, $p$, $\mu$, $\sigma_{ie}$, $\sigma_{pe}$ are given by
\begin{equation}\label{Ms}
M_{s}=\sqrt{\gamma\sigma_{ie}+\frac{(1-p)\sigma_{pe}}{p+\mu(1-\beta_{e}) \sigma_{pe}}},
\end{equation}
\begin{equation}\label{basic_parameters}
\sigma_{ie}=\frac{T_{i}}{T_{e}},\sigma_{pe}=\frac{T_{p}}{T_{e}},
p=\frac{n_{p0}}{N_{0}},\mu=\frac{n_{e0}}{N_{0}},
\end{equation}
and $\beta_{e}$ is the nonthermal parameter associated with the Cairns distributed electrons \cite{cairns1995electrostatic} that determines the proportion of fast energetic electrons. Physically admissible domain of $\beta_{e}$ is $0 \leq \beta_{e}  \leq 4/7 \approx 0.6$.

Here $n_{s0}$ and $T_{s}$ are, respectively, the unperturbed number density and the average temperature of the particles of $s$ - th species, where $s$ = $i$, $e$ and $p$ for ion, electron and positron respectively; $n_{d0}$ is the constant dust number density with $Z_{d}$ is the number of electrons residing on the dust grain surface; $\gamma (= 3)$ is the adiabatic index and $N_{0}=n_{i0}+n_{p0}=n_{e0}+Z_{d}n_{d0}$.

Sardar \textit{et al.} \cite{sardar2016existence} have investigated the existence and stability of the alternative solitary wave solution having a profile different from $\mbox{sech}$ or $\mbox{sech}^{2}$ of the combined MKP - KP equation (\ref{P5_Combined_MKP_KP_equation}) when $L>0$, where $L$ is a function of the parameters.

In the present paper, we have used the same combined MKP-KP equation (\ref{P5_Combined_MKP_KP_equation}) to investigate the existence and stability of the DL solution.

\section{Double layer solutions of the Combined MKP-KP equation}\label{P5_Double_layer_solution}
For DL solution of the combined MKP - KP equation (\ref{P5_Combined_MKP_KP_equation}) propagating along $X$ axis, we make the following change of variables:
\begin{eqnarray}\label{P5_transform variable}
X =\xi -U\tau , \eta^{\prime}=\eta , \zeta^{\prime}=\zeta, \tau^{\prime}=\tau.
\end{eqnarray}
Here, $U$ is the dimensionless velocity (normalized by $C_{D}$, the linearized velocity of the DIA wave in the present plasma system for long-wavelength plane wave perturbation) of the travelling wave moving along $\xi$ - axis, i.e., $U$ is the dimensionless velocity of the wave frame. Under the above changes of the independent variables, the combined MKP-KP equation (\ref{P5_Combined_MKP_KP_equation}) assumes the following form (in which we drop the primes on the independent variables $\eta$, $\zeta$ and $\tau$ to simplify the notations)
\begin{eqnarray}\label{P5_transform form of combined MKP-KP}
\frac{\partial}{\partial X}\Big[-U\phi^{(1)}_{X} + \phi^{(1)}_{\tau}+ AB_{1} \phi^{(1)} \phi^{(1)}_{X} &+& AB_{2}\Big(\phi^{(1)}\Big)^{2}\phi^{(1)}_{X} +\frac{1}{2}A C \phi^{(1)}_{XXX}\Big]\nonumber \\
&+& \frac{1}{2}AD \Big(\phi^{(1)}_{\eta \eta}+\phi^{(1)}_{\zeta \zeta}\Big)=0.
\end{eqnarray}
Now, for the travelling wave solitons of (\ref{P5_transform form of combined MKP-KP}), we set
\begin{eqnarray}\label{P5_phi0X}
\phi ^{(1)} = \phi _{0}(X).
\end{eqnarray}
Substituting (\ref{P5_phi0X}) in (\ref{P5_transform form of combined MKP-KP}), we get
\begin{equation}\label{P5_another transform form of combined MKP-KP}
\frac{d^{2} }{d X^{2}}\Big[-U \phi _{0} + \frac{1}{2} AB_{1}\Big(\phi _{0}\Big)^{2} + \frac{1}{3} AB_{2}\Big(\phi _{0}\Big)^{3}+ \frac{1}{2}A C \frac{d^{2} \phi_{0}}{d X^{2}}\Big] = 0.
\end{equation}
Now, we use the following boundary conditions
\begin{eqnarray}\label{P5_bc1}
\phi_{0}, \frac{d^{n} \phi_{0}}{d X^{n}} \to 0 \mbox{       as       } X \to \infty \mbox{   for   } n=1,2,3,\cdots
\end{eqnarray}
\begin{eqnarray}\label{P5_bc2}
\phi_{0}, \frac{d^{n} \phi_{0}}{d X^{n}} \to 0 \mbox{       as       } X \to -\infty \mbox{   for   } n=1,2,3,\cdots
\end{eqnarray}	
Using the boundary conditions (\ref{P5_bc1}) or (\ref{P5_bc2}), we can write the equation (\ref{P5_another transform form of combined MKP-KP}) as
\begin{equation}\label{P5_for_combined_MKP-KP_solution}
-U \phi _{0} + \frac{1}{2} AB_{1}(\phi _{0})^{2} + \frac{1}{3} AB_{2}(\phi _{0})^{3} + \frac{1}{2}A C \frac{d^{2} \phi_{0}}{d X^{2}} = 0.
\end{equation}
Following the method of Malfliet and Hereman \cite{malfliet1996tanh}, we choose
\begin{equation}\label{P5_solution}
\phi _{0} = a+b ~\mbox{tanh}\frac{X}{W_{1}}
\end{equation}
as a solution of the equation (\ref{P5_for_combined_MKP-KP_solution}). Substituting (\ref{P5_solution}) into (\ref{P5_for_combined_MKP-KP_solution}), we get the following equation:
\begin{equation}\label{exp_R}
\Gamma_{0}+ \Gamma_{1}R+ \Gamma_{2}R^{2}+ \Gamma_{3}R^{3}=0,
\end{equation}
where
\begin{equation}\label{R}
R = \mbox{tanh}\frac{X}{W_{1}}
\end{equation}
and the expressions of $\Gamma_{0}$, $\Gamma_{1}$, $\Gamma_{2}$ and $\Gamma_{3}$
can be written in the following form:
\begin{equation}
\Gamma_{0} =\frac{a}{6} (3aAB_{1}+ 2a^{2}AB_{2}-6U),
\end{equation}
\begin{equation}
\Gamma_{1} = \frac{b}{W_{1}^{2}}(-AC+aAB_{1}W_{1}^{2}+ a^{2}AB_{2}W_{1}^{2}- UW_{1}^{2}),
\end{equation}
\begin{equation}
\Gamma_{2}= \frac{Ab^{2}}{2}(B_{1}+ 2aB_{2}),
\end{equation}
\begin{equation}
\Gamma_{3} = \frac{Ab}{3W_{1}^{2}}(3C+b^{2}B_{2}W_{1}^{2}).
\end{equation}
It is simple to check that the following equation holds good
\begin{equation}
W(1,R, R^{2},R^{3}) = \frac{12}{W_{1}^{6}} \mbox{Sech}^{12}\frac{X}{W_{1}} \neq 0
\end{equation}
for all real $X$, where $W_{1}(1,R, R^{2},R^{3})$ is the Wronskian of the functions $1$, $R$, $R^{2}$ and $R^{3}$. Therefore, the functions $1$, $R$, $R^{2}$ and $R^{3}$ are linearly independent and so, equating the coefficients of different powers of $R$ from both sides of (\ref{exp_R}), we get
\begin{equation}\label{t0}
\frac{a}{6} (3aAB_{1}+ 2a^{2}AB_{2}-6U)=0,
\end{equation}
\begin{equation}\label{t_1}
\frac{b}{W_{1}^{2}}(-AC+aAB_{1}W_{1}^{2}+ a^{2}AB_{2}W_{1}^{2}- UW_{1}^{2})=0,
\end{equation}
\begin{equation}\label{t2}
\frac{Ab^{2}}{2}(B_{1}+ 2aB_{2})=0,
\end{equation}
\begin{equation}\label{t3}
\frac{Ab}{3W_{1}^{2}}(3C+b^{2}B_{2}W_{1}^{2})=0.
\end{equation}
From (\ref{t0}) - (\ref{t3}), we see that if $a=0$ then $b=0$ and conversely, if $b=0$ then $a=0$. Therefore, to get a nontrivial solution of (\ref{t0}) - (\ref{t3}), we assume that $a \neq 0$ and $b \neq 0$. Solving the
equations (\ref{t2}) and (\ref{t3}) for the unknowns $a$ and
$b$, we get
\begin{equation}\label{b}
b^{2}=-\frac{3C}{B_{2}W_{1}^{2}},
\end{equation}
\begin{equation}\label{a}
a=-\frac{B_{1}}{2B_{2}}.
\end{equation}
Using (\ref{b}) and (\ref{a}), from the equation (\ref{t0}) we get
\begin{equation}\label{U1}
U=-\frac{A(4B_{2}C+ B_{1}^{2}W_{1}^{2})}{4B_{2}W_{1}^{2}},
\end{equation}
Using (\ref{b}) and (\ref{a}), from the equation (\ref{t_1}) we get
\begin{equation}\label{U2}
U=-\frac{AB_{1}^{2}}{6B_{2}},
\end{equation}
Now to make a closed and consistent system of equations, the values of $U$ as given by the equations
(\ref{U1}) and (\ref{U2}) must be equal. So, we have the following consistency
condition:
\begin{equation}\label{L=0}
B_{1}^{2}+ 12B_{2}Cp_{1}^{2}=0 \mbox{       with      } p_{1}=\frac{1}{W_{1}}.
\end{equation}
Using (\ref{L=0}), we get the following equation from (\ref{U1}):
\begin{equation}\label{U3}
U=2ACp_{1}^{2}.
\end{equation}
Using (\ref{L=0}), we get the following equation from (\ref{a}):
\begin{equation}\label{b1}
a^{2}=b^{2}.
\end{equation}
Using the boundary condition $\phi_{0} \rightarrow 0$ as $X \rightarrow + \infty$, we get
\begin{equation}\label{bc3}
\lim_{X \rightarrow + \infty}\phi_{0} = (a+b)^{2}=0 \Leftrightarrow b = -a.
\end{equation}
On the other hand, using the boundary condition $\phi_{0} \rightarrow
0$ as $X \rightarrow - \infty$, we get
\begin{equation}\label{bc4}
\lim_{X \rightarrow - \infty}\phi_{0} = (a-b)^{2}=0 \Leftrightarrow b = a,
\end{equation}
Therefore, the DL solution (\ref{P5_solution}) and the condition for its existence can be written as follows:
\begin{equation}\label{P5_phi_final}
 \phi_{0} = a \bigg(1-\lambda  \tanh \frac{X}{W_{1}}\bigg),
\end{equation}
\begin{equation}\label{L1=0}
L = 0,
\end{equation}
where
\begin{equation}\label{a_final}
a = -\frac{B_{1}}{2 B_{2}},
\end{equation}
\begin{equation}\label{L_final}
L = B_{1}^{2}+ 12B_{2}Cp_{1}^{2}.
\end{equation}
Here $\lambda = \pm 1$ and these two values of $\lambda$ give two different DL solutions of the combined MKP-KP equation for the DIA wave corresponding to the the boundary conditions (\ref{P5_bc1}) and (\ref{P5_bc2}) respectively. Specifically, it is impossible to get alternative solitary wave solution of the
combined MKP-KP equation by considering only one of the following two conditions:
\begin{eqnarray}\label{P5_bc1_c_1}
 \phi_{0}, \frac{d^{n} \phi_{0}}{d X^{n}} \to 0 \mbox{       as       } X \to \infty \mbox{   for   } n=1,2,3,\cdots
\end{eqnarray}
\begin{eqnarray}\label{P5_bc2_c_2}
 \phi_{0}, \frac{d^{n} \phi_{0}}{d X^{n}} \to 0 \mbox{       as       } X \to -\infty  \mbox{   for   } n=1,2,3,\cdots
\end{eqnarray}
In fact, using the condition (\ref{P5_bc1_c_1}), we get a Z - type or a S - type  double layer solution of
the  same  combined  MKP-KP  equation  according  to  whether  $ a>0 $ or $ a<0 $ whereas
considering the condition (\ref{P5_bc2_c_2}), we get a S - type or a Z - type  double layer solution of the
same  combined  MKP-KP  equation  according  to  whether  $ a>0 $ or $ a<0 $ . But  both  the
above conditions are necessary to get an alternative solitary wave solution of the combined
MKP-KP equation.

\begin{figure}[t]
	\begin{center}
  \includegraphics{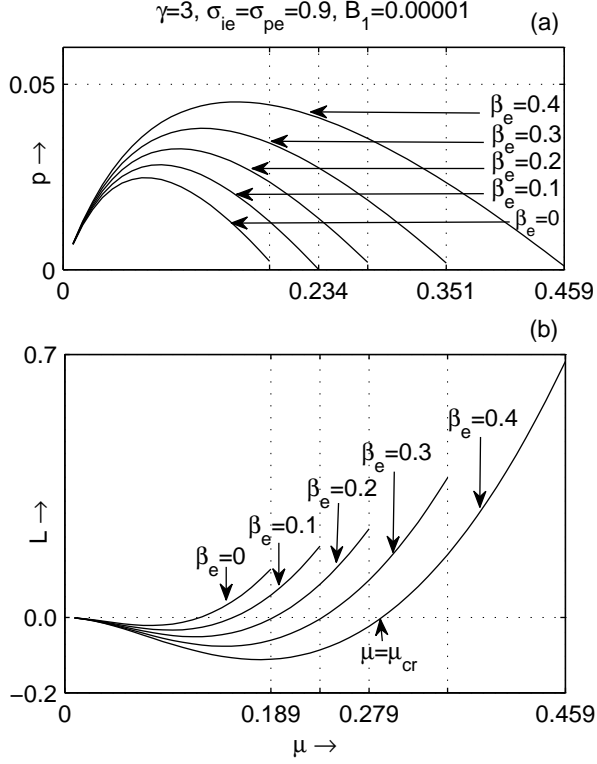}
  \caption{\label{P5_figure4} (a) $p$ is  plotted  against  $\mu$  and (b) $L$ is  plotted  against  $\mu$  for  different  values  of  $\beta_{e}$ with $B_{1}=0.00001$, $\gamma=3$, $\sigma_{ie} = \sigma_{pe}=0.9$ }
  \end{center}
\end{figure}

To discuss the existence of the DL solution (\ref{P5_phi_final}) of the combined MKP - KP equation (\ref{P5_Combined_MKP_KP_equation}), we choose $B_{1}=0.00001$, i.e., we take a small numerical value of $B_{1}$. Now, it is simple to check that $B_{1}$ is a function of $p$, $\mu$ and $\beta_{e}$, i.e., $B_{1}=B_{1}(p, \mu, \beta_{e})$ for fixed values of $\gamma$, $\sigma_{ie}$ and $\sigma_{pe}$. Therefore, $B_{1}(p, \mu, \beta_{e})=0.00001$ gives a functional relation between $\mu$ and $p$ for any fixed value of $\beta_{e}$ within the physically admissible interval of $\beta_{e}$, i.e., $0 \leq \beta_{e} \leq 0.6$. In figure \ref{P5_figure4}(a), this functional relation ($B_{1}(p, \mu, \beta_{e})=0.00001$) between $\mu$ and $p$ is plotted for different values of $\beta_{e}$ with $\gamma=3$, $\sigma_{ie}=0.9$ and $\sigma_{pe}=0.9$. Figure \ref{P5_figure4}(a) shows the existence of a region in the parameter space where $B_{1}=0.00001$.

\begin{figure}[t]
	\begin{center}
  \includegraphics{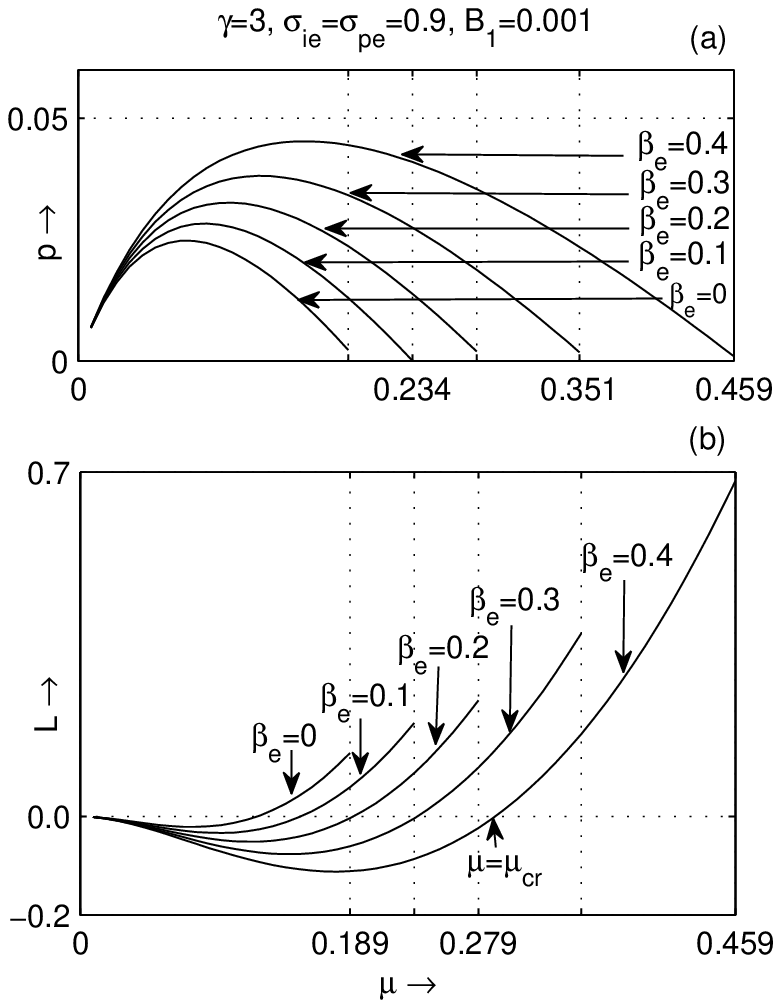}
  \caption{\label{P5_figure2} (a) $p$ is  plotted  against  $\mu$  and (b) $L$ is  plotted  against  $\mu$  for  different  values  of  $\beta_{e}$ with $B_{1}=0.001$, $\gamma=3$, $\sigma_{ie} = \sigma_{pe}=0.9$ }
  \end{center}
\end{figure}

In figure \ref{P5_figure4}(b), $L$ is plotted against $\mu$ for  different  values  of  $\beta_{e}$ with $\gamma=3$, $\sigma_{ie} = \sigma_{pe}=0.9$ when $B_{1}=0.00001$, i.e., in figure \ref{P5_figure4}(b), L is plotted against $\mu$ along the each curve of figure \ref{P5_figure4}(a). Figure \ref{P5_figure4}(b) clearly shows that there exists a value $\mu_{cr}$ of $\mu$ such that $L>0$ or $L<0$ according to whether $\mu > \mu_{cr}$ or $\mu < \mu_{cr}$ and $L=0$ at $\mu = \mu_{cr}$. Obviously, in the small neighbourhood of $\mu = \mu_{cr}$, $L$ is close to zero.  Again, for each $\beta_{e}$ lying within a fixed interval, there exists a value $\mu_{cr}$ of $\mu$ such that $L=0$. So, for $L=0$, we can use the DL solution (\ref{P5_phi_final}) as the solution of the combined MKP-KP equation (\ref{P5_Combined_MKP_KP_equation}). In this connection, we want to mention the following point of  Sardar \textit{et al.} \cite{sardar2016existence} Sardar \textit{et al.} \cite{sardar2016existence} reported that the alternative solitary wave solution of the combined MKP - KP equation (\ref{P5_Combined_MKP_KP_equation}) cannot describe the nonlinear behaviour of DIA waves when $L=0$ or $L \approx O(\epsilon)$ and further investigation is necessary when $L=0$ or $L \approx O(\epsilon)$.

\begin{figure}[t]
	\begin{center}
  \includegraphics{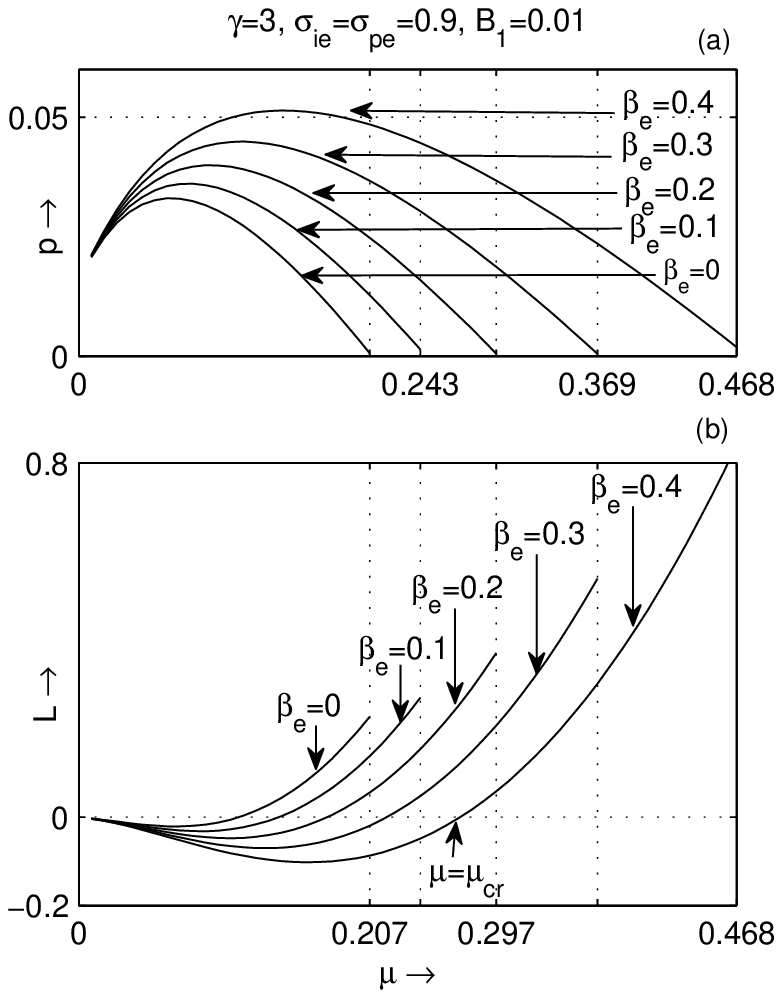}
  \caption{\label{P5_figure1} (a) $p$ is  plotted  against  $\mu$  and (b) $L$ is  plotted  against  $\mu$  for  different  values  of  $\beta_{e}$ with $B_{1}=0.01$, $\gamma=3$, $\sigma_{ie} = \sigma_{pe}=0.9$ }
  \end{center}
\end{figure}

Similarly, figure \ref{P5_figure2}(a) shows the existence of a region in the parameter space where $B_{1}=0.001$ and figure \ref{P5_figure2}(b) shows the existence of the points $\mu_{cr}$ of $\mu$ such that $L=0$ at $\mu = \mu_{cr}$. Figure \ref{P5_figure1}(a) shows the existence of a region in the parameter space where $B_{1}=0.01$ and figure \ref{P5_figure1}(b) shows the existence of the points $\mu_{cr}$ of $\mu$ such that $L=0$ at $\mu = \mu_{cr}$. Sardar \textit{et al.} \cite{sardar2016existence} have already drawn the existence of a region in the parameter space where $B_{1}=0.0001$ in figure 2(a) whereas in figure 2(b), they have drawn the the existence of the points $\mu_{cr}$ of $\mu$ such that $L=0$ at $\mu = \mu_{cr}$ for $B_{1}=0.0001$. From the above mentioned figures, we can conclude that we can draw the existence of a region in the parameter space for any given small value of $B_{1}$ and for this value of $B_{1}$, we can easily find the existence of the point $\mu_{cr}$ of $\mu$ such that $L=0$ at $\mu = \mu_{cr}$ for any fixed value of $\beta_{e}$ and for different values of $\beta_{e}$, one can get different $\mu_{cr}$ such that $L=0$ at $\mu = \mu_{cr}$. So, in the present paper, we are able to give a consistent solution of the combined MKP - KP equation (\ref{P5_Combined_MKP_KP_equation}) when $L=0$.

The solution (\ref{P5_phi_final}) is the steady state solution of the combined MKP - KP equation (\ref{P5_Combined_MKP_KP_equation}) along the $x$ - axis. This solution is same as the DL solution of the combined MKdV - KdV equation corresponding to the combined MKP - KP equation (\ref{P5_Combined_MKP_KP_equation}), i.e., the steady state solitary wave solution of the combined MKdV - KdV equation
\begin{equation}\label{P5_Combined_MKdV_KdV_equation}
\phi^{(1)}_{\tau} + AB_{1}
\phi^{(1)} \phi^{(1)}_{\xi}  + AB_{2} (\phi^{(1)})^{2} \phi^{(1)}_{\xi} + \frac{1}{2} A C \phi^{(1)}_{\xi \xi \xi}
= 0.
\end{equation}
is exactly same as the equation (\ref{P5_phi_final}). In the present paper, we have considered the lowest order transverse stability of the DL solution of the combined MKdV - KdV equation (\ref{P5_Combined_MKdV_KdV_equation}) using the three-dimensional combined MKP - KP equation (\ref{P5_Combined_MKP_KP_equation}). In fact, to study the transverse stability of the DL solution of the combined MKdV - KdV equation (\ref{P5_Combined_MKdV_KdV_equation}), we have considered the three-dimensional combined MKP - KP equation (\ref{P5_Combined_MKP_KP_equation}) by taking the weak dependence of the spatial coordinates perpendicular to the direction of propagation of the wave.
\section{Stability analysis}\label{P5_Stability_Double_layer}
In section \ref{P5_Double_layer_solution}, we have seen that the equation (\ref{P5_phi_final}) gives two different DL solutions of the combined
MKP-KP equation (\ref{P5_Combined_MKP_KP_equation}) for two different values of $\lambda$, viz., $\lambda=+1$ and $\lambda=-1$. In this section, we have considered the stability analysis of the double  layer solution (\ref{P5_phi_final}) for $\lambda=+1$. Following the same procedure, one can easily analyse the stability of the DL solution (\ref{P5_phi_final}) for $\lambda=-1$.

To analyze the stability of the DL solution (\ref{P5_phi_final}) of the equation (\ref{P5_transform form of combined MKP-KP}) by the multiple-scale perturbation expansion method of Allen and Rowlands \cite{allen1993determination,allen1995stability},  we decompose $\phi^{(1)}$ as
\begin{eqnarray}\label{P5_decompose phi}
\phi^{(1)}=\phi_{0}(X)+ q(X, \eta, \zeta, \tau),
\end{eqnarray}
where $\phi_{0}(X)$ is the DL solution (\ref{P5_phi_final}) of the equation (\ref{P5_transform form of combined MKP-KP}) and $q(X, \eta, \zeta, \tau)$ is the perturbed part of $\phi^{(1)}$. Substituting (\ref{P5_decompose phi}) into (\ref{P5_transform form of combined MKP-KP}) and linearizing this equation with respect to $q$ we get the following equation:
\begin{equation}\label{P5_equation_of_q_Linear}
q_{X\tau}+ \Big(M_{1}q \Big)_{XX}+ \frac{1}{2}AD (q_{\eta \eta}+ q_{\zeta \zeta})=0,
\end{equation}
where
\begin{equation}\label{M1_equation}
M_{1}=-U + AB_{1} \phi_{0} + AB_{2} \phi_{0}^{2}+ \frac{1}{2} AC \frac{\partial^{2}}{\partial X^{2}}
\end{equation}
and  we have used the following notations:
\begin{equation}
q_{X\tau} = \frac{\partial^{2} q}{\partial X \partial \tau}~,~\Big(M_{1}q \Big)_{XX} = \frac{\partial^{2} }{\partial X^{2}}\Big( M_{1}q \Big).
\end{equation}
Now, for long-wavelength plane-wave perturbation along a direction  having direction cosines ( $l$, $m$, $n$), we set
\begin{equation}\label{P5_exp of q}
q(X, \eta, \zeta, \tau) = \overline{q}(X)e^{i\{k(l X+m \eta+n \zeta)-\omega\tau\}},
\end{equation}
where $k$ is small and $l^{2}+m^{2}+n^{2}=1$.

Substituting the expression of $q(=q(X, \eta, \zeta, \tau))$ as given in (\ref{P5_exp of q}) into the equation (\ref{P5_equation_of_q_Linear}), we get the following equation of $\overline{q}$:
\begin{eqnarray}\label{P5_equation_of_q_bar}
&& (M_{1}\overline{q})_{XX}-i \omega \overline{q}_{X} +kl\Big\{\omega \overline{q}+2i(M_{1}\overline{q})_{X}+iAC \overline{q}_{XXX}\Big\}\nonumber \\
&& - k^{2}l^{2}\Big\{M_{1}\overline{q}+\frac{5}{2}AC \overline{q}_{XX}+AD\frac{m^{2}+n^{2}}{2l^{2}}\overline{q}\Big\} -k^{3}l^{3}\Big\{2iAC\overline{q}_{X}\Big\}+k^{4}l^{4}\Big\{\frac{1}{2}AC\overline{q}\Big\}=0,\nonumber \\
\end{eqnarray}
Following the multiple-scale perturbation expansion method of Allen and Rowlands \cite{allen1993determination,allen1995stability}, we expand $\overline{q}(X)$ and $\omega$ as
\begin{equation}\label{P5_expansion of_{q}(X)}
\overline{q}(X) = \sum_{j=0}^{\infty}k^{j}q^{(j)}(X,X_{1},X_{2},X_{3},...),
\end{equation}
\begin{equation}\label{P5_omega}
\omega = \sum_{j=0}^{\infty}k^{j}\omega^{(j)},
\end{equation}
where $\omega^{(0)}=0$,  $X_{j}=k^{j}X$, $j=0,1,2,3...,$ and each $q^{(j)}(=q^{(j)}(X,X_{1},X_{2},X_{3},...))$ is a function of $X,X_{1},X_{2},X_{3},\cdots$ . It is important to note that $X_{0}=X$.

Finally, substituting (\ref{P5_expansion of_{q}(X)}) and (\ref{P5_omega}) into the equation (\ref{P5_equation_of_q_bar}) and then equating the coefficients of different powers of $k$ on the both sides of the resulting equation, we get the following sequence of equations:
\begin{equation}\label{P5_sequence_of_eq}
\frac{\partial}{\partial X}(M_{1} q^{(j)})=Q^{(j)},
\end{equation}
where
\begin{equation}\label{P5_Q^{j}_equation}
Q^{(j)}=\int_{\infty}^{X}{R^{(j)}} dX,
\end{equation}
and the expression of $R^{(j)}$ for $j$ = $0$ and $1$ are given in the following equations:
\begin{eqnarray}
R^{(0)} &=& 0,\label{R0}\\
R^{(1)} &=& i\omega^{(1)}q^{(0)}_{0}- iAClq^{(0)}_{000} - ACq^{(0)}_{0001} -2il[M_{1}q^{(0)}]_{0}- 2 [M_{1}q^{(0)}]_{01}.\label{P5_R1}
\end{eqnarray}
Here, we have used the following notations:
\begin{eqnarray}\label{qjr}
q^{(j)}_{r} = \frac{\partial q^{(j)}}{\partial X_{r}},~
q^{(j)}_{rs} = \frac{\partial^{2} q^{(j)}}{\partial X_{r}\partial X_{s}},~
q^{(j)}_{rst} = \frac{\partial^{3} q^{(j)}}{\partial X_{r}\partial X_{s}\partial X_{t}}\nonumber
\end{eqnarray}
\begin{eqnarray}\label{qjry}
q^{(j)}_{rsty} = \frac{\partial^{4} q^{(j)}}{\partial X_{r}\partial X_{s}\partial X_{t}\partial X_{y}},~
[M_{1}q^{(j)}]_{rs}=\frac{\partial^{2} (M_{1}q^{(j)})}{\partial X_{r}\partial X_{s}}\nonumber
\end{eqnarray}

Assuming that $q^{(j)}$ and its derivative up to third order vanish as $X \rightarrow \infty$, the general solution of (\ref{P5_sequence_of_eq}) can be written as
\begin{eqnarray}\label{P5_general solution}
q^{(j)} &=& A^{(j)}_{1}f+A^{(j)}_{2}f\int\frac{1}{f^2}dX+A^{(j)}_{3}f\int\frac{\phi_{0}}{f^{2}}dX+ \frac{2}{AC} f\int\frac{\int (f\int Q^{(j)}dX)dX}{f^{2}}dX,\nonumber \\
\end{eqnarray}
where $\displaystyle f=\frac{d\varphi_{0}}{dX}$, $\phi _{0}$ is given by (\ref{P5_phi_final}) for $\lambda=1$ and $A_{1}^{(j)}, A_{2}^{(j)}, A_{3}^{(j)}$ are all arbitrary functions of $X_{1}, X_{2}, X_{3}, \cdots $.

Using MATHEMATICA \cite{wolfram1999mathematica}, the solution (\ref{P5_general solution}) for $\lambda=+1$ can be put in the following form:
\begin{eqnarray}\label{P5_qj_general}
q^{(j)} &=& A_{1}^{(j)} f +\frac{3W_{1}^{2}}{8a^{2}} \Big( A_{2}^{(j)}
+ a A_{3}^{(j)} \Big) fX + \frac{3W_{1}^{2}}{8a^{2}} \Big( A_{2}^{(j)}
+ a A_{3}^{(j)} \Big) \phi_{0}\nonumber \\
&+& \frac{W_{1}^{2}}{8a} \Big( A_{2}^{(j)}
+ a A_{3}^{(j)} \Big) e^{-\frac{2X}{W_{1}}} - \frac{W_{1}^{2}}{4a}{A_{2}^{(j)}}\frac{1}{S^{2}}- \frac{W_{1}^{2}}{4a} \Big(A_{2}^{(j)} + a A_{3}^{(j)}\Big)\nonumber\\
&+& \frac{2}{AC} f \int\frac{\int (f \int Q^{(j)}dX)dX}{f^{2}}dX,
\end{eqnarray}
where
\begin{equation}
S=\mbox{sech}\frac{X}{W_{1}}.
\end{equation}
\subsection{Zeroth order equation}

As $R^{(0)}=0$, the solution (\ref{P5_qj_general}) of the equation (\ref{P5_sequence_of_eq}) for $j=0$ can be written as follows :
\begin{eqnarray}\label{P5_q0}
q^{(0)} &=& A_{1}^{(0)} f +\frac{3W_{1}^{2}}{8a^{2}} \Big( A_{2}^{(0)} + a A_{3}^{(0)} \Big) fX + \frac{3W_{1}^{2}}{8a^{2}} \Big( A_{2}^{(0)} + a A_{3}^{(0)} \Big) \phi_{0}\nonumber \\
&+& \frac{W_{1}^{2}}{8a} \Big( A_{2}^{(0)} + a A_{3}^{(0)} \Big) e^{-\frac{2X}{W_{1}}} - \frac{W_{1}^{2}}{4a}{A_{2}^{(0)}}\frac{1}{S^{2}}- \frac{W_{1}^{2}}{4a} \Big(A_{2}^{(0)} + a A_{3}^{(0)}\Big).
\end{eqnarray}
We note that each term of $q^{(0)}$ except the fifth term  $\displaystyle \Big[- \frac{W_{1}^{2}}{4a}{A_{2}^{(0)}}\frac{1}{S^{2}}\Big]$ is bounded at $X = +\infty$. Therefore, to make $q^{(0)}$ bounded at $X = +\infty$, the coefficient of
$\frac{1}{S^{2}}$ in the expression of $q^{(0)}$ must be identically equal to zero and consequently we get
\begin{equation}\label{P5_A20}
- \frac{W_{1}^{2}}{4a}{A_{2}^{(0)}}=0.
\end{equation}
Now, we see that each term of $q^{(0)}$ except the last term $\displaystyle \Big[- \frac{W_{1}^{2}}{4a} \Big(A_{2}^{(0)} + a A_{3}^{(0)}\Big)\Big]$ approaches zero as $X\rightarrow +\infty$. So, to make $q^{(0)}$ consistent with the condition $q^{(0)}\rightarrow 0$ as $X\rightarrow
+\infty$, we must have
\begin{equation}\label{P5_A20_A30}
- \frac{W_{1}^{2}}{4a} \Big(A_{2}^{(0)} + a A_{3}^{(0)}\Big)=0.
\end{equation}
From equations (\ref{P5_A20}) and (\ref{P5_A20_A30}), we get,
\begin{equation}\label{P5_A20_A30_final}
A_{2}^{(0)}=A_{3}^{(0)}=0.
\end{equation}
Therefore $q^{(0)}$ assumes the following form:
\begin{equation}\label{P5_q0_final}
q^{(0)} = A_{1}^{(0)}f = A_{1}^{(0)} \frac{d\phi_{0}}{dX}.
\end{equation}

\subsection{First order equation}
Using the equations (\ref{P5_R1}) , (\ref{P5_q0_final}) and MATHEMATICA \cite{wolfram1999mathematica}, the solution (\ref{P5_qj_general}) of the differential equation (\ref{P5_sequence_of_eq}) for $j=1$ can be put in the following form:
\begin{eqnarray}\label{P5_q1}
q^{(1)} &=& B_{1}^{(1)} f + \Big\{ \frac{3W_{1}^{2}}{8a^{2}} \Big( A_{2}^{(1)} + a A_{3}^{(1)} \Big) + i A_{1}^{(0)}s_{1}- \frac{\partial A_{1}^{(0)} }{\partial X_{1}} \Big\} fX \nonumber \\
&+& \Big\{ \frac{3W_{1}^{2}}{8a^{2}} \Big( A_{2}^{(1)} + a A_{3}^{(1)}\Big)+ i A_{1}^{(0)} \frac{\omega^{(1)}}{2U}  \Big\} \phi_{0}\nonumber \\ &+& \Big\{\frac{W_{1}^{2}}{8a} \Big( A_{2}^{(1)} + a A_{3}^{(1)} \Big)++ i A_{1}^{(0)} \frac{a\omega^{(1)}}{2U}  \Big\} e^{-\frac{2X}{W_{1}}} \nonumber\\
&-& \frac{W_{1}^{2}}{4a}{A_{2}^{(1)}}\frac{1}{S^{2}}- \frac{W_{1}^{2}}{4a} \Big(A_{2}^{(1)} + a A_{3}^{(1)}\Big),
\end{eqnarray}
where
\begin{equation}\label{B11}
B_{1}^{(1)}= A_{1}^{(1)}+ i A_{1}^{(0)} \frac{3W_{1}}{8U} \omega^{(1)},
\end{equation}
\begin{equation}\label{P5_s1}
s_{1} = \frac{ \omega^{(1)}- 2lU}{2U}.
\end{equation}
We note that each term of $q^{(1)}$ except the fifth term  $\displaystyle \Big[- \frac{W_{1}^{2}}{4a}{A_{2}^{(1)}}\frac{1}{S^{2}}\Big]$ is bounded at $X = +\infty$. Therefore, to make $q^{(1)}$ bounded at $X = +\infty$, the coefficient of
$\frac{1}{S^{2}}$ in the expression of $q^{(1)}$ must be identically equal to zero and consequently we get
\begin{equation}\label{P5_A12}
- \frac{W_{1}^{2}}{4a}{A_{2}^{(1)}}=0.
\end{equation}
Now, we see that each term of $q^{(1)}$ except the last term $\displaystyle \Big[- \frac{W_{1}^{2}}{4a} \Big(A_{2}^{(1)} + a A_{3}^{(1)}\Big)\Big]$ approaches zero as $X\rightarrow +\infty$. So, to make $q^{(1)}$ consistent with the condition $q^{(1)}\rightarrow 0$ as $X\rightarrow
+\infty$, we must have
\begin{equation}\label{P5_A12_A13}
- \frac{W_{1}^{2}}{4a} \Big(A_{2}^{(1)} + a A_{3}^{(1)}\Big)=0.
\end{equation}
From equations (\ref{P5_A12}) and (\ref{P5_A12_A13}), we get,
\begin{equation}\label{P5_A12_A13_final}
A_{2}^{(1)}=A_{3}^{(1)}=0.
\end{equation}
Therefore $q^{(1)}$ assumes the following form:
\begin{equation}\label{P5_q1_final}
q^{(1)} = B_{1}^{(1)} f+ \Big( i A_{1}^{(0)}s_{1}- \frac{\partial A_{1}^{(0)} }{\partial X_{1}} \Big) fX  +  i A_{1}^{(0)} \frac{\omega^{(1)}}{2U} \phi_{0}+  i A_{1}^{(0)} \frac{a\omega^{(1)}}{2U} e^{-\frac{2X}{W_{1}}}.
\end{equation}
Now, we can remove the first term in the above expression of $q^{(1)}$
as this type of term has already been included in the lowest order
term $q^{(0)}$. The second term in the above expression of
$q^{(1)}$ is known as ghost secular term and this term can be removed by choosing
\begin{equation}\label{A01X1}
\frac{\partial A_{1}^{(0)}}{\partial X_{1}}= i A_{1}^{(0)}s_{1}.
\end{equation}
Therefore, the equation (\ref{P5_q1_final}) can be written as
\begin{eqnarray}\label{P5_q1_final_1}
q^{(1)} &=&   i A_{1}^{(0)} \frac{\omega^{(1)}}{2U} \phi_{0}+  i A_{1}^{(0)} \frac{a\omega^{(1)}}{2U} e^{-\frac{2X}{W_{1}}}.
\end{eqnarray}
Now we see that the first term of $q^{(1)}$ is bounded at $X=\pm \infty$ but the second term of $q^{(1)}$ is not bounded at $X=- \infty$ because of the presence of the term $e^{-\frac{2X}{W_{1}}}$. To make $q^{(1)}$ bounded at $X=- \infty$, We must have
\begin{equation}\label{P5_w_1}
iA_{1}^{(0)}\frac{a\omega^{(1)}}{2U} =0.
\end{equation}
The equation (\ref{P5_w_1}) gives the following expression for $\omega^{(1)}$:
\begin{eqnarray}\label{First_order_final}
\omega^{(1)}=0.
\end{eqnarray}
Equation (\ref{First_order_final}) shows that $\omega^{(1)}$ is real and consequently, the DL solution (\ref{P5_phi_final}) for $\lambda=1$ is stable at the order $k$, where $k$ is the wave number of the perturbation.

\section{Conclusions}\label{P5_conclusions}
 In the present paper, we have considered the problem of existence and stability of the DL solution of the combined MKP-KP equation. Analytically, we have proved that the DL solution having a profile of the form (\ref{P5_phi_final}) of the combined MKP-KP equation exists if $L = 0 $. The form of the double layer solution as given by the equation (\ref{P5_phi_final}) also suggests that there are two types of double layer solutions corresponding to two different values of $\lambda$ ($\lambda=+1$ and $\lambda = -1$). We have stated earlier that these double layer solutions of the combined MKP-KP equation are exactly same as those of the combined MKdV-KdV equation but here we consider the combined MKP-KP equation to study the stability of the double layer solutions of the corresponding combined MKdV-KdV equation. Finally, we have found that double layers are stable up to the lowest order of the wave number. We have the following differences between the paper of Sardar \textit{et al.} \cite{sardar2016existence} and the present paper:
\begin{itemize}
	\item (1) In the paper of Sardar \textit{et al.} \cite{sardar2016existence}, the method of Malfliet and Hereman \cite{malfliet1996tanh} has been used to find the alternative solitary wave solution of the combined MKP-KP equation having profile different from $\mbox{sech}^{2/r}$ for any strictly positive real value of $r$ whereas in the present paper, following the tan-hyperbolic method of Malfliet and Hereman \cite{malfliet1996tanh}, we have found the double layer solution of the same combined MKP-KP equation.
	\item (2) In the paper of Sardar \textit{et al.} \cite{sardar2016existence}, they have used the following boundary conditions:
\begin{equation}\label{P5_bc1_c_another}
\phi_{0}, \frac{d^{n} \phi_{0}}{d X^{n}} \to 0 \mbox{  as } |X| \to \infty  \nonumber 
\end{equation}
$\mbox{  for  } n=1,2,3,\cdots $ to get the alternative solitary wave solution of the combined MKP-KP equation whereas in present paper, we have used the following conditions:
\textbf{either} $\displaystyle \phi_{0}, \frac{d^{n} \phi_{0}}{d X^{n}} \to 0 \mbox{  as } X \to \infty$ \textbf{or} $\displaystyle \phi_{0}, \frac{d^{n} \phi_{0}}{d X^{n}} \to 0 \mbox{  as } X \to -\infty$ $\mbox{  for  } n=1,2,3,\cdots $ to get a double layer solution of the same combined MKP-KP equation.
	\item (3) In the paper of Sardar \textit{et al.} \cite{sardar2016existence}, they got the following alternative solitary wave solution of the combined MKP-KP equation:
\begin{equation}\label{phi1}
\phi_{0}=12Cp_{1}^{2}\frac{sech[2p_{1}X]}{B_{1}sech[2p_{1}X]+\lambda \sqrt{L}} ,\nonumber
\end{equation}
where $\lambda=\pm1$, $L=B_{1}^{2}+12B_{2}Cp_{1}^{2}>0$ and $p_{1}=\frac{1}{W_{1}}$, whereas in the present paper, we get the following double layer solution of the same combined MKP-KP equation:
\begin{equation*}\label{P5_phi_final_c}
 \phi_{0} = a \bigg(1-\lambda  \tanh \frac{X}{W_{1}}\bigg),\nonumber
\end{equation*}
where $a = -\frac{B_{1}}{2 B_{2}}$.

Here $\lambda =1$ if we use the conditions $\phi_{0}, \frac{d^{n} \phi_{0}}{d X^{n}} \to 0 \mbox{  as  } X \to \infty$ and $\lambda =-1$ if we use the conditions $\phi_{0}, \frac{d^{n} \phi_{0}}{d X^{n}} \to 0 \mbox{       as       } X \to -\infty$ $\mbox{   for   } n=1,2,3,\cdots$ .
	\item (4) In the paper of Sardar \textit{et al.} \cite{sardar2016existence}, it has been shown that the alternative solitary wave solution of the combined MKP-KP equation exists only when $L>0$ whereas in the present paper, it has been shown that the double layer solution of the same combined MKP-KP equation exists when $L=0$.
	\item (5) In the paper of Sardar \textit{et al.} \cite{sardar2016existence}, it has been shown that the alternative solitary wave solution of the combined MKP-KP equation converges to the solitary wave solution of MKP equation when $B_{1}\rightarrow 0$ whereas in the present paper, we see that the double layer solution of the same combined MKP-KP equation does not converge to the solitary wave solution of the MKP equation when $B_{1}\rightarrow0$ and it is impossible to get any double layer solution of the MKP equation.
\end{itemize}

Regarding  stability  analysis,  we  want  to mention  that  we  have  used  the  small-$k$ perturbation expansion method of Rowlands and Infeld \cite{rowlands1969stability,infeld1972stability,infeld1973stability,zakharov1974instability,infeld1985self} to analyse the lowest order stability of the alternative solitary wave solution of the combined MKP-KP equation whereas we have used the multiple-scale perturbation expansion method of Allen and Rowlands \cite{allen1993determination} to analyse the lowest order stability of the double layer solution of the combined MKP-KP equation. In fact, one  cannot  use  the  small-$k$  perturbation  expansion  method  of  Rowlands  and  Infeld \cite{rowlands1969stability,infeld1972stability,infeld1973stability,zakharov1974instability,infeld1985self} to analyse  the  lowest  order  stability  of  double  layer  solution  of  the  combined  MKP-KP equation.


\providecommand{\noopsort}[1]{}\providecommand{\singleletter}[1]{#1}%

\end{document}